\documentstyle[aps,epsfig]{revtex}

\begin{document}

\preprint{
\vbox{
\hbox{July 2000}
\hbox{...}
}}

\title{The Temperature Dependence of the QCD Running Coupling}

\author{F. M. Steffens}

\address{   NFC - FCBEE - Universidade Presbiteriana Mackenzie,
            Rua da Consola\c{c}\~ao 930, 01302-907,
            S\~ao Paulo, SP, Brazil \\
            IFT - UNESP,
            Rua Pamplona 145, 01405-900,
            S\~ao Paulo, SP, Brazil}

\maketitle

\begin{abstract}
We study the running of the QCD coupling with the momentum squared
($Q^2$) and the temperature scales in the high temperature limit
($T > T_{c}$), using a mass dependent renormalization scheme to
build the Renormalization Group Equations. The approach used
guaranty gauge invariance, through the use of the Hard Thermal
Loop approximation, and independence of the vertex chosen to
renormalize the coupling. In general, the dependence of the
coupling with the temperature is not logarithmical, although in
the region $Q^2 \sim T^2$ the logarithm approximation is
reasonable. Finally, as known from Debye screening, color charge
is screened in the coupling. The number of flavors, however, is
anti-screened.

\end{abstract}


\vspace*{1cm}

One of the important questions in perturbative QCD at finite
temperature $T$ is what is the temperature dependence of the
strong coupling constant, $\alpha_s (T)$. According to an early
study by Collins and Perry \cite{cp}, where arguments from the
Renormalization Group Equations (RGE) were used, temperature
replaces the momentum as the running scale, implying that
$\alpha_s (T)$ decreases as a logarithm  as $T$ increases. This
conclusion is based on general considerations, where no specific
calculations of the vertices and self-energies were done. Later,
more rigorous treatments based on the RGE were developed
\cite{kapusta,matsumoto}, with applications to perturbative QCD
\cite{kapusta,nakkagawa}. However, it was soon realized that the
resulting coupling was strongly dependent on both the gauge
\cite{kobes,chaichian,sasaki} and vertex chosen to renormalize it
\cite{nakkagawa,fujimoto,baier}. In any case, some of the
calculations did not result in the logarithmical dependence of the
coupling in the temperature. Recently, in the context of a
semiclassical background field method, it has also been calculated
\cite{schneider} that $\alpha_s$ has an inverse logarithmical
dependence in the temperature, where reference to the earlier
calculations based on the RGE was said to substantiate the result.
Moreover, applications to the perturbative calculations of energy
density, pressure, etc, using the logarithmical dependence are
frequently used in the literature \cite{peshier,weise,meisinger}.

In the early 1990´s the problem of gauge invariance of two, three
and four points Green functions was solved with the introduction
of the Hard Thermal Loops (HTL) \cite{braaten}. In particular, it
was shown that the dominant, gauge invariant, contributions to the
N-point functions have a $T^2$ dependence, and that they satisfy
abelian type Ward identities \cite{braaten,taylor}. From the gluon
self-energy in the HTL, it is possible to build an effective
coupling at finite temperature, which goes with the inverse of the
Debye mass and thus does not have a logarithmic dependence on the
temperature \cite{book}. However, there has been no extensive
calculation of the RGE for the ultra-violet regime of the running
coupling using the dominant, gauge invariant, results from the
HTL. Rigorously, there is no extra ultra-violet divergence induced
by thermal effects, meaning that the renormalization of QCD at
$T=0$ is sufficient to make QCD at $T\neq 0$ finite. Still, we can
redefine the theory by a finite renormalization, relating
parameters like mass and coupling constants at different
temperatures with the aid of temperature dependent renormalization
constants, with appropriate renormalization conditions
\cite{matsumoto}. The purpose of this letter is to calculate the
running of the QCD coupling with the momentum scale and with the
temperature using the vertices and self-energies calculated from
the HTL. Although it is an immediate application, we think that
given the discrepancies found in the literature, this is an useful
exercise with an impact in all the perturbative QCD calculations
at finite temperature.

We will treat the mass scale introduced in the system by the
temperature in the same way quark masses are handled in the $T=0$
theory \cite{gp76}, in a mass dependent renormalization scheme. In
this case, the general form for the renormalized n-points Green
function will be:
\begin{equation}
\Gamma_n = \Gamma_n (k, g^2, k^2/\mu^2, T^2/\mu^2),
\end{equation}
where $\mu$ is the renormalization scale, $k$ is the external
momentum and $g$ the renormalized coupling at the scale $\mu$.
Using the standard approach of the RGE, i.e., taking the
derivative of  the unrenormalized Green function with respect to
$\mu$, one gets:
\begin{equation}
\left[\mu\frac{\partial}{\partial\mu} +
\beta\left(g^2,\frac{T^2}{\mu^2}\right) \frac{\partial}{\partial
g^2} +
\gamma_T\left(g^2,\frac{T^2}{\mu^2}\right)T\frac{\partial}{\partial
T} + \gamma_n\left(g^2,\frac{T^2}{\mu^2}\right)\right] \Gamma_n =
0,
\end{equation}
where $\gamma_n\left(g^2,\frac{T}{\mu}\right)$ is the anomalous
dimension of the $n$ fields appearing in $\Gamma_n$ and
\begin{eqnarray}
\beta\left(g^2,\frac{T^2}{\mu^2}\right) &=& \mu\frac{d g^2}{d\mu},
\nonumber \\* \gamma_T\left(g^2,\frac{T^2}{\mu^2}\right)
&=&\frac{\mu}{T}\frac{dT}{d\mu},
\label{e3}
\end{eqnarray}
the beta function and the temperature anomalous dimension,
respectively. The temperature dependence in the beta function and
anomalous dimensions indicates that we are working in a mass
dependent renormalization scheme. The relation between the bare
and renormalized coupling is given by:

\begin{equation}
\alpha_s (\mu^2, T^2)= Z^{-1}_\alpha (\mu^2,T^2) \alpha_0,
\label{e4}
\end{equation}
where $Z_\alpha$ is the renormalization constant of the coupling,
$\alpha_0 = g_0^2/4\pi$ is the bare coupling, and $\alpha_s =
g^2/4\pi$ is the coupling at the subtraction point $\mu$ and
temperature $T$.

If we want a meaningful result for the running coupling, any
vertex should be equally good to renormalize it. If we use the
triple gluon vertex, then

\begin{equation}
Z_\alpha^{-1}(\mu^2,T^2/\Lambda_T^2) =
\frac{Z_3^3(\mu^2,T^2/\Lambda_T^2)}{Z_1^2(\mu^2,T^2/\Lambda_T^2)},
\label{e5}
\end{equation}
where $Z_3$ renormalizes the gluon field while $Z_1$ renormalizes
the triple gluon vertex. We start with the renormalization of the
gluon field. To include temperature in the renormalization
constants and simultaneously preserve the Slavnov-Taylor
identities, we will work in the framework of the HTL resummation
program, where the dominant terms are known to be gauge invariant.
A direct computation of the transverse part of the gluon
self-energy at one loop gives \cite{braaten2,brandt}:

\begin{equation}
\Pi_T^{(1)} = - \left(N_c + \frac{n_f}{2}\right) \frac{g_0^2
T^2}{12 |\vec k|^2} \left[\frac{k_0}{|\vec k|} ln\left(\frac{k_0 +
|\vec k|}{k_0 - |\vec k|}\right) - 2 \frac{k_0^2}{k^2}\right],
\label{e6}
\end{equation}
where $k$ is the four momentum of the external gluon. Only the
dominant term in the high temperature expansion was written
because of its gauge invariance. One can see that Eq. (\ref{e6})
vanishes for $k_0 << |\vec k|$, and that it is reduced to

\begin{equation}
\Pi_T^{(1)} = \frac{\alpha_0}{\pi} \left(N_c +
\frac{n_f}{2}\right) \frac{2 \pi^2}{3} \frac{T^2}{\Lambda_T^2}
\label{e7}
\end{equation}
for $k_0 >> |\vec k|$, and $\Lambda_T$ is a mass scale
($\Lambda_T^2 \equiv |\vec k|k_0 $).

We will be working in a Momentum Subtraction Scheme (MOM), which
is suitable for a mass dependent renormalization scheme as in the
present case, with the temperature replacing the mass. Although a
MOM scheme usually breaks the Slavnov-Taylor identities, the
deviation among the couplings defined through different vertices
is vanishingly small \cite{smilga}. As a renormalization
condition, we impose that the thermal part of the renormalized
gluon self-energy, calculated at the subtraction point $\mu^2$, be
given by:

\begin{equation}
\Pi_{T,R}^{(1)} = \frac{\alpha_0}{\pi} \left(N_c +
\frac{n_f}{2}\right) \frac{2 \pi^2}{3} \frac{T^2}{\mu^2}.
\label{e8}
\end{equation}
Including the $T=0$ part, we then calculate the renormalization
constant $Z_3$:

\begin{equation}
Z_3 = 1 + Z_3^{(1)} = 1 +  \frac{\alpha_0}{\pi} \left(N_c +
\frac{n_f}{2}\right) \frac{2 \pi^2}{3} \frac{T^2}{\mu^2}
\frac{\Lambda_T^2 - \mu^2}{\Lambda_T^2} + \frac{\alpha_0}{12 \pi}
\left[\frac{13 - 3a}{2} N_c - 2 n_f\right]ln
\frac{\Lambda^2}{\mu^2}, \label{e9}
\end{equation}
where $a$ is the gauge parameter. We now need the one loop
correction to the triple gluon vertex, which is a known quantity
\cite{book}: it has the same functional form of Eq. (\ref{e7}).
Imposing again that the temperature dependent part of the
renormalized vertex be equal to the bare one, with the replacement
of $\Lambda_T \rightarrow \mu$, we calculate the triple gluon
vertex renormalization constant to be:

\begin{equation}
Z_1 = 1 + Z_1^{(1)} = 1 +  \frac{\alpha_0}{\pi} \left(N_c +
\frac{n_f}{2}\right) \frac{2 \pi^2}{3} \frac{T^2}{\mu^2}
\frac{\Lambda_T^2 - \mu^2}{\Lambda_T^2} + \frac{\alpha_0}{12 \pi}
\left[\frac{17 - 9a}{4}N_c - 2 n_f\right]ln
\frac{\Lambda^2}{\mu^2}. \label{e10}
\end{equation}
Using Eqs. (\ref{e9}) and (\ref{e10}) in (\ref{e5}), we get:

\begin{equation}
Z_\alpha^{-1} = 1 +  \frac{\alpha_0}{\pi} \left(N_c +
\frac{n_f}{2}\right) \frac{2 \pi^2}{3} \frac{T^2}{\mu^2}
\frac{\Lambda_T^2 - \mu^2}{\Lambda_T^2} + \frac{\alpha_0}{12 \pi}
\left[11 N_c - 2 n_f\right]ln \frac{\Lambda^2}{\mu^2}.
\label{e11}
\end{equation}
Notice that the temperature dependence, unlike the momentum scale,
is not logarithmical. Before we proceed to calculate the running
of the coupling, we make a few remarks on the calculation of
$Z_\alpha^{-1}$ using other vertices. For instance, if we use the
ghost-gluon vertex, then:

\begin{equation}
Z_\alpha^{-1} = \frac{Z_3 \tilde{Z_3}^2}{\tilde{Z_1}^2},
\label{e12}
\end{equation}
where $\tilde{Z_3}$ is the ghost field renormalization constant
and $\tilde{Z_1}$ is the gluon-ghost vertex renormalization
constant. It happens that in the HTL approximation the external
momenta in the numerators of the loop integrals are disregarded,
as the main contribution comes from the internal momenta of order
$T$, with $T$ taken to be large. Compared to the gluon
self-energy, the ghost self-energy does not have enough powers of
internal momenta in the numerator to produce the leading, $T^2$,
behavior in the temperature. The same happens to the ghost vertex,
something that can be immediately inferred from the abelian Ward
identities relating the ghost vertex to its self-energy. Hence,
the thermal parts of $\tilde{Z_3}$ and $\tilde{Z_1}$ will be
sub-leading to $Z_3$ in the temperature. In such case, the thermal
part of $Z_\alpha$ will be dominated by $Z_3$ and we recover Eq.
(\ref{e11}). Finally, we could have used the quark-gluon vertex to
calculate $Z_\alpha$:

\begin{equation}
Z_\alpha^{-1} = \frac{Z_3 Z_2^2}{Z_{1F}^2},
\label{e13}
\end{equation}
where $Z_2$ is the fermion field renormalization constant while
$Z_{1F}$ is the quark-gluon vertex renormalization constant. The
HTL calculation for the dominant part of the quark self-energy and
the quark-gluon vertex at one loop are known \cite{book}. Their
$T^2$ dependence are the same. Imposing renormalization conditions
of the type of Eq. (\ref{e8}) for the calculation of $Z_2$ and
$Z_{1F}$, it follows that their $T^2$ dependence are the same,
cancelling in the ratio of Eq. (\ref{e13}). As before, $Z_3$ will
dominate the temperature dependence of $Z_\alpha$, and we once
again recover Eq. (\ref{e11}).

The calculation of the $\beta$ function (including the $T=0$ and
the $T \neq 0$ parts), for a fixed temperature but an arbitrary
renormalization point is now straightforward. Using Eqs.
(\ref{e3}), (\ref{e4}), and (\ref{e11}), we have:

\begin{equation}
\mu\frac{d\alpha_s}{d\mu} =
\frac{\alpha_s^2}{\pi}\left[-\frac{11}{6}N_c + \frac{2}{6} n_f -
\frac{2\pi^2}{3}\left(N_c + \frac{n_f}{2}\right) \mu
\frac{d}{d\mu}\frac{T^2}{\mu^2} \right], \label{e14}
\end{equation}
to order $\alpha_s^2$. This is the only RGE for $\alpha_s$ because
there is only one renormalization scale for both the $T=0$ and $T
\neq 0$ parts. The solution of Eq. (\ref{e14}) is:

\begin{equation}
\alpha_s(Q^2,T^2) = \frac{\alpha_s(Q_0^2,T^2)} {1 +
\frac{\alpha_s(Q_0^2,T^2)}{4\pi}\left[\left(\frac{11}{3}N_c -
\frac{2}{3}n_f \right)ln \left(\frac{Q^2}{Q_0^2}\right) +
\frac{8\pi^2}{3} \left(N_c + \frac{n_f}{2}\right)
\left(\frac{T^2}{Q^2} - \frac{T^2}{Q_0^2}\right) \right]}.
\label{e15}
\end{equation}
It is helpful to rewrite Eq. (\ref{e15}) in the same format of the
$T=0$ theory. To this end, we define effective, scale dependent,
numbers of colors and flavors:

\begin{equation}
N_c^{eff}(T^2,Q^2,Q_0^2) = \left[1 + \frac{8\pi^2}{11}
\frac{T^2/Q^2 - T^2/Q_0^2}{ln \left(\frac{Q^2}{Q_0^2}\right)}
\right] N_c, \label{e16}
\end{equation}

\begin{equation}
n_f^{eff}(T^2,Q^2,Q_0^2) = \left[1 - 2 \pi^2 \frac{T^2/Q^2 -
T^2/Q_0^2} {ln \left(\frac{Q^2}{Q_0^2}\right)} \right] n_f.
\label{e17}
\end{equation}
With the help of Eqs. (\ref{e16}) and (\ref{e17}), the expression
for the running coupling is written as:

\begin{equation}
\alpha_s(Q^2, T^2) = \frac{\alpha_s(Q_0^2, T^2)} {1 +
\frac{\alpha_s(Q_0^2, T^2)}{4\pi}\left[\frac{11}{3}N_c^{eff} -
\frac{2}{3}n_f^{eff}\right] ln \left(\frac{Q^2}{Q_0^2}\right)}.
\label{e18}
\end{equation}
Equations (\ref{e16}) and (\ref{e17}) tell us that for fixed
$Q_0^2$ and $T^2$, with $Q^2 \rightarrow 0$, the effective number
of color decreases, $N_c^{eff} < N_c$, while the effective number
of flavors increases: color charge is screened, and the number of
flavors are anti-screened. The same argument applies when keeping
$Q^2$ and $Q_0^2$ fixed, while taking $T^2$ going to infinity.  To
quantify these assertions, we show in Fig. 1 the $Q^2$ dependence
of $n_f^{eff}$ and $N_c^{eff}$ for $T^2 = 1\; GeV^2$, $Q_0^2 =
m_Z^2$, and $N_c = n_f = 3$. As expected, $N_c^{eff}$ decreases
while $n_f^{eff}$ increases as $Q^2$ decreases from $Q_0^2$ to
$T^2$.

\begin{figure}[htb]
\centerline{\epsfig{file=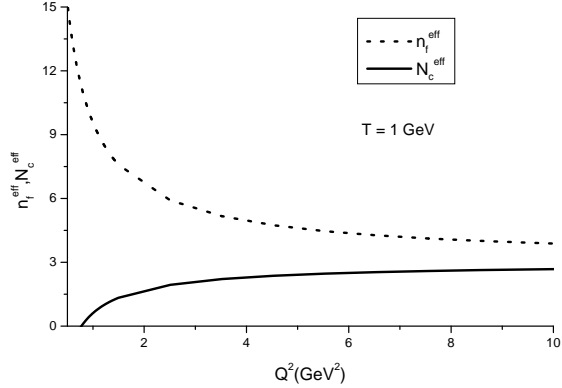,height=6cm}} \caption{The
effective number of colors and flavors at $T = 1\; GeV$ as a
function of $Q^2$. For large values of $Q^2$, both tend to their
values in the $T= 0\; GeV$ theory.} \label{fig1}
\end{figure}

In Fig. 2 we show the behavior of $\alpha_s$ for 3 values of the
temperature. For $\alpha_s(Q_0^2, T^2)$, we use the experimental
value measured at $m_Z \approx 91 \; GeV$ and at zero temperature
\cite{pdg}. That is, we assume that at such high values of the
virtuality of the probe, temperatures of the order of $1 \; GeV$
are not relevant. This seems to be the case as, in this region,
$N_c^{eff} \rightarrow N_c$ and $n_f^{eff}\rightarrow n_f$. At
$T=0 \; GeV$ we have, as usual, the coupling growing rapidly for
$Q^2 < 10 \; GeV^2$. However, at $T= 0.5$ and $ 1 \; GeV$,
$\alpha_s(Q^2,T^2)$ starts to change its behavior in the region
around $Q^2 = 20\; GeV^2$. Instead of the rapid growth observed at
the $T = 0$ case, for finite $T$ there is first an almost $Q^2$
independence of the coupling, and then it {\it decreases} with
$Q^2$, showing the color screening, and flavor anti-screening, in
action. In fact, we observe that $\alpha_s$ changes its
qualitative behavior (from a divergence to finite values), for
small values of $Q^2$, in the region around $T \sim 0.2 \; GeV$,
although any conclusion based on a perturbative RGE analysis for
such low values of $T$ and $Q^2$ should be taken with extreme
caution. In any case, in general the running coupling does not
have a logarithm dependence on the temperature. But if we consider
the particular region where $Q^2 \sim T^2$, taking $Q_0^2
>> T^2$, we get from Eq. (\ref{e16}-\ref{e18}) that:

\begin{equation}
\alpha_s(Q^2 \sim T^2, T^2) \rightarrow \alpha_s(T^2) \simeq
\frac{12\pi}{(11N_c - 2n_f)ln
\left(\frac{T^2}{\Lambda_{QCD}^2}\right)}. \label{e19}
\end{equation}

\begin{figure}
\centerline{\epsfig{file=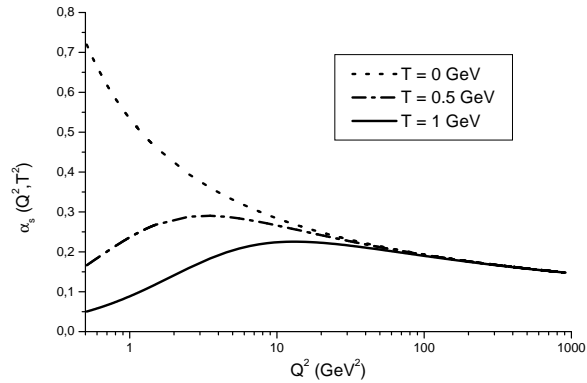,height=6cm}} \caption{The
strong coupling constant as a function of $Q^2$ calculated for 3
differerent values of the temperature.} \label{fig2}
\end{figure}

In summary, we have presented a RGE calculation of the running
coupling at finite temperature QCD, introducing the temperature
scale as the mass scale is introduced in the $T=0$ theory, in a
mass dependent renormalization scheme. We restricted the
calculation to the dominant, gauge invariant, HTL terms. As a
result, the coupling is shown to be independent of the vertex used
to renormalize it. As is well known from Debye screening, the
number of colors are screened by the temperature. The number of
flavors, however, is anti-screened. Also, although in general the
temperature dependence of the coupling is not logarithmic, in some
especial cases, where $Q^2 \sim T^2$, the logarithmic dependence
can be a good approximation.

\acknowledgements

I would like to thank F. T. Brandt for helpful discussions on
thermal field theory, and the support of the Special Research
Centre for the Subatomic Structure of Matter at the University of
Adelaide during the initial stages of this work. This work was
supported by FAPESP (03/10754-0), CNPq (308932/2003-0) and
MackPesquisa .


{\bf 339},  310 (1990).
\bibitem{taylor} J. Frenkel and J. C. Taylor,
Nucl. Phys. B {\bf 334}, 199 (1990).

\bibitem{book} Michel Le Bellac,
in {\it Thermal Field Theory}, edited by P. V. Landshoff, D. R.
Nelson, D. W. Sciana and S. Weiberg (Cambridge University Press,
Cambridge, 1996)

\bibitem{gp76} H. Georgi and H. D. Politzer, Phys. Rev. D {\bf 14},
1829 (1976).

\bibitem{braaten2} E. Braaten and R. D. Pisarski,
Phys. Rev. D {\bf 42}, 2156 (1990).

\bibitem{brandt} F. T. Brandt and J. Frenkel, Phys. Rev. D {\bf 56}, 2453 (1997);
F. T. Brandt, J. Frenkel and F. R. Machado,
Phys. Rev. D {\bf 61}, 125014 (2000).


\bibitem{smilga} Andrei Smilga, {\it Lectures on Quantum
Chromodynamics}, World Scientific 2001.

\bibitem{pascual} P. Pascual and R. Tarrach,
in {\it QCD: Renormalization for the Practitioner},
edited by H. Araki {\it et al.} (Springer Verlag, Berlin -
Heidelberg, 1984).

\bibitem{pdg} Particle Data Group,
Phys. Rev. D {\bf 54}, 1 (1996).



\references

\bibitem{cp} J. C. Collins and M. J. Perry, Phys. Rev. Lett. {\bf
34}, 1353 (1975).

\bibitem{kapusta} J. I. Kapusta, Phys. Rev. D {\bf 20}, 989 (1979).

\bibitem{matsumoto} H. Matsumoto, Y. Nakano and Y. Umezawa, Phys.
Rev. D {\bf 29}, 1116 (1984).

\bibitem{nakkagawa} H. Nakkagawa and A. Ni\'egawa, Phys. Lett. B
{\bf 193}, 263 (1987); H. Nakkagawa, A. Ni\'egawa
and H. Yokota, Phys. Rev. D {\bf 38}, 2566 (1988).

\bibitem{kobes} P. Elmfors and R. Kobes, Phys. Rev. D {\bf 51}, 774 (1995).

\bibitem{chaichian} M. Chaichian and M. Hayashi, Acta. Phys. Polon
{\bf 27}, 1703 (1996).

\bibitem{sasaki} K. Sasaki, Nucl. Phys. B {\bf 490}, 472 (1997).

\bibitem{fujimoto} Y. Fujimoto and H. Yamada,
Phys. Lett. B {\bf 200}, 167 (1988).

\bibitem{baier} R.Baier, B. Pire and D. Schiff,
Phys. Lett. B {\bf 238}, 367 (1990).

\bibitem{schneider} R. A. Schneider, Phys. Rev. D {\bf 67}, 057901
(2003); Phys. Rev. D {\bf 66}, 036003 (2002).

\bibitem{peshier} A. Peshier et al., Phys. Rev. D {\bf 54}, 2399
(1996).

\bibitem{weise} R. A. Schneider and W. Weise, Phys. Rev. C {\bf
64}, 055201 (2001).

\bibitem{meisinger} P. N. Meisinger, M. C. Ogilvie and T. R.
Miller, Phys. Lett. B {\bf 585}, 149 (2004).

\bibitem{braaten} E. Braaten and R. D. Pisarski,
Phys. Rev. Lett. {\bf 64}, 1338 (1990);
Nucl. Phys. B {\bf 337}, 569 (1990);
\end{document}